\documentclass{pasa}%

\usepackage{graphicx}

\title[Spin direction asymmetry in DESI Legacy Survey]{Large-scale asymmetry in galaxy spin directions: evidence from the Southern hemisphere}

\author[Lior Shamir]{Lior Shamir
\affil{Kansas State University \\ Manhattan, KS 66506}%
}%

\usepackage{aas_macros}
\usepackage{hyperref} 
\hypersetup{colorlinks,citecolor=blue,linkcolor=blue,urlcolor=blue}

\hypersetup{draft}
\begin{document}

\begin{frontmatter}
\maketitle

\begin{abstract}

Recent observations using several different telescopes and sky surveys showed patterns of asymmetry in the distribution of galaxies by their spin directions as observed from Earth. These studies were done with data imaged from the Northern hemisphere, showing excellent agreement between different telescopes and different analysis methods. Here, data from the DESI Legacy Survey was used. The initial dataset contains $\sim2.2\cdot10^7$ galaxy images, reduced to $\sim8.1\cdot10^5$ galaxies annotated by their spin direction using a symmetric algorithm. That makes it not just the first analysis of its kind in which the majority of the galaxies are in the Southern hemisphere, but also by far the largest dataset used for this purpose to date. The results show strong agreement between opposite parts of the sky, such that the asymmetry in one part of the sky is similar to the inverse asymmetry in the corresponding part of the sky in the opposite hemisphere. %Separating the Southern hemisphere into two opposite regions shows a statistically significant higher number of clockwise galaxies in one part of the Southern hemisphere, and a statistically significant higher number of counterclockwise galaxies in the opposite part of the Southern hemisphere. 
Fitting the distribution of galaxy spin directions to cosine dependence shows a dipole axis with probability of 4.66$\sigma$. Interestingly, the location of the most likely axis is within close proximity to the CMB Cold Spot. The profile of the distribution is nearly identical to the asymmetry profile of the distribution identified in Pan-STARRS, and it is within 1$\sigma$ difference from the distribution profile in SDSS and HST. All four telescopes show similar large-scale profile of asymmetry.

% large datasets of galaxies showed non-random distribution of the spin directions of spiral galaxies, even when the galaxies are too far from each other to have gravitational interaction. Here, a dataset of $\sim8.7\cdot10^3$ spiral galaxies imaged by Hubble Space Telescope is used to test and profile a possible asymmetry between galaxy spin directions. The asymmetry between galaxies with opposite spin directions is compared to the asymmetry of galaxies from the Sloan Digital Sky Survey. The two datasets contain different galaxies at different redshift ranges, and each dataset was annotated using a different annotation method. The results show that both datasets show a similar asymmetry in the COSMOS field, which is covered by both telescopes. Fitting the asymmetry of the galaxies to cosine dependence shows a dipole axis with probabilities of $\sim2.8\sigma$ and $\sim7.38\sigma$ in HST and SDSS, respectively. The most likely dipole axis identified in the HST galaxies is at $(\alpha=78^o,\delta=47^o)$, and is well within the $1\sigma$ error range compared to the location of the most likely dipole axis in the SDSS galaxies with $z>0.15$, identified at $(\alpha=71^o,\delta=61^o)$. 
\end{abstract}

\begin{keywords}
Galaxy: general -- galaxies: spiral
\end{keywords}
\end{frontmatter}

\section{INTRODUCTION}
\label{introduction}

Autonomous digital sky surveys that collect very large astronomical databases have allowed to address research questions that their studying was not possible in the pre-information era. One of these questions is the large-scale distribution of galaxy spin directions as observed from Earth. While the initial assumption would be that the spin directions of galaxies are randomly distributed at a large scale, there is no valid proof to that assumption. In fact, multiple experiments have shown that the distribution as seen from Earth might not be random \citep{longo2007cosmic,longo2011detection,shamir2012handedness,shamir2019large,shamir2020pasa,shamir2020patterns,shamir2020large,lee2019galaxy,lee2019mysterious,shamir2021particles}, and the profile of the distribution could exhibit a large-scale dipole axis \citep{shamir2020pasa,shamir2020patterns,shamir2021particles}. When normalized for the redshift distribution, different telescopes show very similar profiles of asymmetry, and the directions of the most likely axes agree within statistical error \citep{shamir2020pasa,shamir2020patterns}. 

The observation of a large-scale axis around which the Universe is oriented has been proposed in the past through observations of cosmic microwave background (CMB), with consistent data from the Cosmic Background Explorer (COBE), Wilkinson Microwave Anisotropy Probe (WMAP) and Planck \citep{mariano2013cmb,land2005examination,ade2014planck,santos2015influence}.

% Unlike elliptical galaxies, the visual appearance of a spiral galaxy depends on the perspective of the observer, making spiral galaxies a unique type of extra-galactic objects. 

Early attempts to study the distribution of spin directions of spiral galaxies were based primarily by manual annotating a large number of galaxy images  \citep{iye1991catalog,land2008galaxy,longo2011detection}. Since human annotation is too slow to analyze large datasets, and can also be affected by the bias of the human perception \citep{hayes2017nature}, handling larger datasets in a systematic manner requires automatic analysis of the galaxies \citep{shamir2012handedness,shamir2013color,shamir2016asymmetry,shamir2019large,shamir2020pasa,shamir2020patterns}. The application of the automatic analysis methods to different datasets acquired by different telescopes showed very similar profiles of distribution in data from Sloan Digital Sky Survey (SDSS), the Panoramic Survey Telescope and Rapid Response System (Pan-STARRS), and the Hubble Space Telescope \citep{shamir2020pasa,shamir2020patterns}.

Non-random distribution of spin directions of galaxies has been also observed in cosmic filaments \citep{tempel2013evidence,tempel2013galaxy}. %although that spin is different than the two-headed alignment, where the spins show no preference compared to the direction of the alignment. 
Alignment in spin directions associated with the large-scale structure was also observed with quasars \citep{hutsemekers2014alignment} and smaller sets of spiral galaxies \citep{lee2019mysterious}. In addition to the observational studies, dark matter simulations have also shown links between spin direction and the large-scale structure \citep{zhang2009spin,libeskind2013velocity,libeskind2014universal}. The strength of the correlation has been associated with stellar mass and the color of the galaxies \citep{wang2018spin}. These links were associated with halo formation \citep{wang2017general}, leading to the contention that the spin in the halo progenitors is linked to the large-scale structure of the early universe \citep{wang2018build}. % It should be mentioned that the spin direction of a galaxy might not be necessarily the same as the spin direction of the dark matter halo, as it has been proposed that a galaxy can also spin in a different direction than its host halo \citep{wang2018spin}.

The numerous studies with different approaches, messengers, telescopes, and datasets that suggest links between the large-scale structure and the direction in which extra-galactic objects spin reinforce the studying of this question using larger datasets. Here, the distribution of spin directions of spiral galaxies is studied using a very large number of galaxies from the Southern hemisphere.

\section{DATA}
\label{data}

The dataset of spiral galaxies was taken from the DESI Legacy Survey \citep{dey2019overview}, which is a combination of data collected by the Dark Energy Camera (DECam), the Beijing-Arizona Sky Survey (BASS), and the Mayall z-band Legacy Survey (MzLS). The imaging is calibrated to provide a dataset of nearly uniform depth \citep{dey2019overview}.

The list of objects was determined by using all South bricks of Data Release (DR) 8 of the DESI Legacy Survey. Objects with g magnitude smaller than 19.5 and identified as exponential disks (``EXP''), de Vaucouleurs ${r}^{1/4}$ profiles (``DEV''), or round exponential galaxies (`'REX'') in DESI Legacy Survey DR8 were selected. That selection provided a list of objects identified as extended objects, but excluded objects that may be embedded in other extended objects. %Objects embedded in other extended objects might lead to an error in the identification of the galaxy center, which can lead to incorrect annotation of the spin direction. 
Objects identified as galaxies but are part of other extended objects such as H II regions can still exist in the dataset, and it is virtually impossible to inspect the entire dataset by eye and remove such potential objects. If such object is located on the arm of a galaxy, it might be wrongly identified as the center of that galaxy, and can lead to incorrect annotation. As will be explained in Section~\ref{conclusion}, if such error indeed exists, it is expected to impact clockwise and counterclockwise similarly, and therefore cannot lead to asymmetry in the dataset.

The images were retrieved through the {\it cutout} API of the DESI Legacy Survey server. The images were downloaded as 256$\times$256 JPEG images, scaled by the Petrosian radius to ensure the galaxy fits in the frame. The total number of images retrieved from the Legacy Survey server is 22,987,246. Downloading such a high number of images required a substantial amount of time. The first image was retrieved on June 4th, 2020, and the process continued consistently until March 4th, 2021, with just short breaks due to power outages or system updates of the computer that was used to download the images. All images were downloaded by using the same server located at Kansas State University campus to avoid any possible differences between the way different computers download and store images.

Annotation of the spin directions of a large number of galaxies is highly impractical to perform manually due to the labor involved in the process. Therefore, the annotation of the galaxies requires automation. A valid automatic process for this task should be mathematically symmetric, to avoid any possible systematic bias. Machine learning is commonly used to annotate images automatically in multiple domains, including astronomy. However, machine learning, and especially convolutional neural networks (CNNs), are normally based on complex non-intuitive data-driven rules, and almost always have a small but non-negligible error. Because these rules are determined automatically during the training process, there is no guarantee that these rules are symmetric. These rules are sensitive to the training data, and in fact are unlikely to be fully symmetric. To eliminate the possibility that such bias affects the results, the algorithm needs to be deterministic and mathematically symmetric.

To avoid the possibility of systematic bias, the fully symmetric model-driven Ganalyzer algorithm was used \citep{shamir2011ganalyzer}. Ganalyzer first converts each galaxy image into its radial intensity plot. The radial intensity plot is the transformation of the original galaxy image such that the value of the pixel $(x,y)$ in the radial intensity plot is the median value of the 5$\times$5 pixels around $(O_x+\sin(\theta) \cdot r,O_y-\cos(\theta)\cdot r)$ in the original image, where {\it r} is the radial distance, $\theta$ is the polar angle, and $O_x$, $O_y$ are the X and Y pixel coordinates of the center of the galaxy.

After the galaxy image is converted into its radial intensity plot, a peak detection algorithm \citep{morhavc2000identification} is applied to the horizontal lines of the radial intensity plot to identify the peaks in each row. Because the arms of the galaxy have brighter pixels than other pixels at the same distance from the galaxy center, the peaks in the radial intensity plot are the galaxy arms in the original image. Since the arm of a spiral galaxy is curved, the peaks are expected to form a line towards the direction in which the arm spins.  When applying a linear regression to that line, the sign of the regression determines the direction towards which the arm is curved, and consequently the direction towards which the galaxy spins \citep{shamir2011ganalyzer,shamir2017photometric,shamir2017colour,shamir2017large,shamir2019large,shamir2020patterns}.

\begin{figure}[h]
\centering
\includegraphics[scale=0.35]{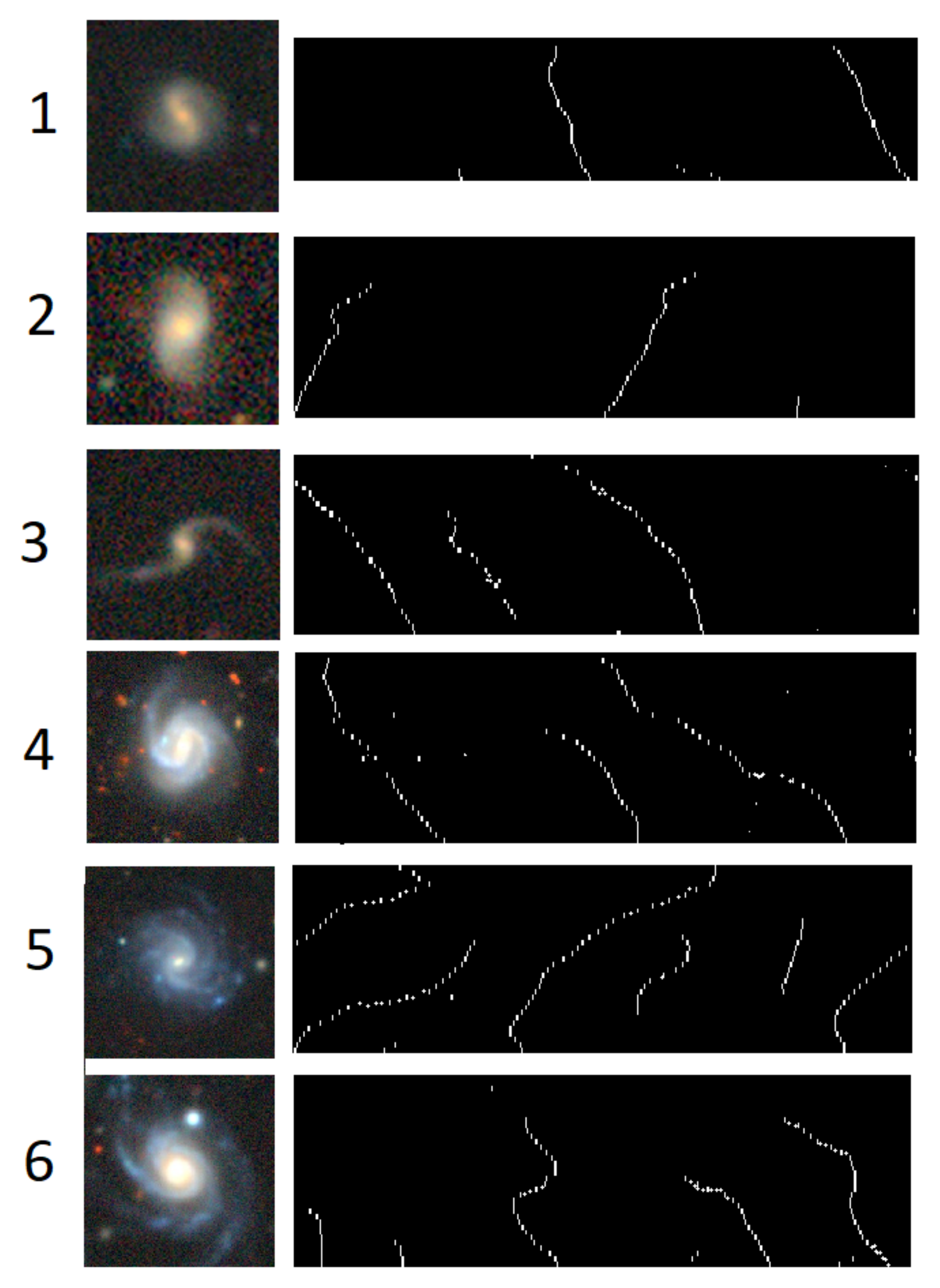}
\caption{Examples of the peaks of the radial intensity plots of different galaxy images. The direction of the lines generated by the peaks identifies the curves of the galaxy arms, and therefore can be used to determine the spin directions. The algorithm is fully symmetric, and is not based on complex non-intuitive data-driven rules commonly used in machine learning.}
\label{radial_intensity_plot}
\end{figure}

Figure~\ref{radial_intensity_plot} shows examples of galaxy images and the peaks of the radial intensity plot transformation of each galaxy. The first two images show a simple galaxy structure with two arms, and therefore the peaks of the radial intensity plots have two lines. Each line corresponds to an arm of a galaxy. The direction of the lines reflects the spin direction of the galaxies. Galaxies 3 and 4 are galaxies with three arms, as reflected by the lines created by the peaks of the radial intensity plots. Galaxies 5 and 6 have more complex morphology, but still can be identified by a higher number of peaks that drift towards the direction that allows to identify the spin direction of the galaxy.

It is clear that not all galaxies in the initial dataset are spiral, and not all spiral galaxies provide clear details that can allow to identify their spin direction. To remove galaxies that their spin direction cannot be identified with high certainty, only galaxies that have at least 30 identified peaks aligned in lines are considered as galaxies with identifiable spin directions. Additionally, the number of peaks that shift in one direction should be at least three times the number of peaks shifting towards the opposite direction \citep{shamir2020patterns}. If that condition is not satisfied, the galaxy is not annotated, and being rejected from the analysis. 

The requirement for the peaks of the radial intensity plot to shift to one direction at least three times more than the other direction can handle situations in which the arms are tilted in a non-uniform manner. Galaxy 6 in Figure~\ref{radial_intensity_plot} is an example of a galaxy with arms tilted in non-uniform directions. The number of peaks shifting to the right (counterclockwise) is 55, while the number of peaks shifting to the left is 18. That reflects the stronger counterclockwise winding as seen in the image. If the number of peaks shifting to the left was higher, the galaxy would not be annotated at all, would be assumed inconclusive, and consequently rejected from the analysis. It is reasonable to assume that many galaxies with identifiable spin directions are rejected from the analysis, reducing the size of the usable data. However, the key requirement of the algorithm is that the rejection of galaxies is done in a symmetric manner, and with no preference to galaxies of a certain spin direction. As discussed in Section~\ref{conclusion} and shown theoretically and empirically in \citep{shamir2021particles}, Ganalzyer is mathematically symmetric. Also, experiments that were done by adding a large number of artificial incorrectly classified galaxies showed minor impact on the results of the analysis when they were distributed evenly, while even 1\% artificial systematic error led to very strong signal that peaked at the celestial pole \citep{shamir2021particles}. Additional discussion about the impact of different types of possible errors can be found in Section~\ref{conclusion}. More detailed information about the annotation algorithm can be found in \citep{shamir2011ganalyzer,shamir2012handedness,hoehn2014characteristics,dojcsak2014quantitative,shamir2021particles,shamir2020patterns,shamir2020pasa}.

After applying the algorithm, 836,451 galaxies were assigned with identifiable spin directions. In some cases digital sky survey can identify objects that are part of the same galaxy as separate galaxies. To avoid having the same galaxy more than once in the dataset, objects that have another object within less than 0.01$^o$ were removed. After removing these objects, 807,898 galaxies were left in the dataset. To test the consistency of the classification, 200 galaxies annotated as spinning clockwise and 200 galaxies annotated as spinning counterclockwise were inspected manually. None of the galaxies determined by the algorithm to be spinning clockwise was visually spinning counterclockwise, and none of the galaxies that according to the algorithm were spinning counterclockwise was by manual inspection spinning clockwise. While that simple test could not guarantee that no galaxy is annotated incorrectly, it can indicate that the galaxies are separated into two groups such that one has more galaxies spinning clockwise and the other has more galaxies spinning counterclockwise. Incorrect annotations are expected to be distributed evenly between clockwise and counterclockwise galaxies, as discussed in Section~\ref{conclusion}.

To ensure that the process of annotation is consistent, all galaxy images were analyzed by the exact same algorithm, exact same code, exact same computer, and exact same processor. That was done to avoid a situation in which different computers or even different processors analyze different galaxies, and could lead to differences in the way each galaxy is analyzed. Annotation of the initial dataset of over 2.2$\cdot10^7$ galaxies required 107 days using a single Intel Xeon processor at 2.8 Ghz. To reduce the response time of the experiment, the process of galaxy annotation started before all galaxies were fully downloaded from the DESI Legacy Survey server.

Like many other digital sky surveys, galaxies in the Legacy Surveys are not distributed uniformly across all right ascension and declination ranges. Tables~\ref{ra_distribution} and~\ref{dec_distribution} show the number of galaxies in different right ascension and declination ranges, respectively. As Table~\ref{dec_distribution} shows, the dataset also includes bricks with positive declination. All galaxies in Table~\ref{dec_distribution} were used, and therefore the dataset was not limited to merely galaxies with negative declination. While the dataset is not made of purely galaxies from the Southern hemisphere, most galaxies have far lower declination compared to data collected by telescopes located in the Northern hemisphere such as Pan-STARRS or SDSS. Pan-STARRS and SDSS also cover some of the Southern sky, but do not go below declination of -30$^o$, and most of their footprint is in positive declination. Table~\ref{z_distribution} shows the redshift distribution of the galaxies. The Legacy Survey galaxies do not have redshift, and therefore the distribution of the redshift was measured by using a subset of 17,027 galaxies that had redshift in the 2dF redshift survey \citep{cole20052df}. % The mean redshift of the galaxies was 0.124. % and DR16 of SDSS. 

% \begin{table}
% \caption{The number of galaxies in different right ascension ranges.}
% \label{ra_distribution}
% \centering
% \begin{tabular}{lc}
% \hline
% \hline
% RA range & \# galaxies  \\
% (degrees) &                  \\
% \hline
% 0-60     &  289,311  \\  
% 60-120  &  101,220  \\
% 120-180 &  112,502 \\ 
% 180-240 & 117,011 \\	
% 240-300 & 39,092 \\
% 300-360 & 148,762  \\
% \hline
% \hline
% \end{tabular}
% \end{table}

\begin{table}
\caption{The number of galaxies in different right ascension ranges.}
\label{ra_distribution}
\centering
\begin{tabular}{lc}
\hline
\hline
RA range & \# galaxies  \\
(degrees) &                  \\
\hline
0-30     &  155,628  \\  
30-60  &  133,683 \\
60-90 &   80,134 \\ 
90-120 & 21,086 \\	
120-150 & 52,842 \\
150-180 &  59,660 \\
180-210 &  58,899 \\
210-240 & 58,112 \\
240-270 & 36,490 \\
270-300 & 2,602 \\
300-330 & 64,869 \\
330-360 & 83,893 \\
\hline
\hline
\end{tabular}
\end{table}

\begin{table}
\caption{The number of galaxies in different declination ranges.}
\label{dec_distribution}
\centering
\begin{tabular}{lc}
\hline
\hline
Declination         & \# galaxies  \\
range (degrees) &                  \\
\hline
-70 - -50    &  81,355  \\  
-50 - -30   &   123,972  \\
-30 - -10   &  121,656 \\ 
-10 - +10  &  236,740 \\	
+10 - + 30 & 203,562  \\
+30 - + 50 &  40,613 \\
\hline
\hline
\end{tabular}
\end{table}

\begin{table}
\caption{The number of galaxies in different z ranges. The distribution is determined by a subset of 17,027 galaxies that had redshift in the 2dF redshift survey.}
\label{z_distribution}
\centering
\begin{tabular}{lc}
\hline
\hline
z       & \# galaxies  \\
\hline
0-0.05    &  2,089  \\  
0.05-0.1   & 5,487    \\
0.1-0.15   &  4,226  \\ 
0.15 - 0.2  &  1,927 \\	
0.2-0.25 & 784 \\
0.25 - 0.3 &   621 \\
$>$0.3 &  1,893 \\
\hline
\hline
\end{tabular}
\end{table}

\section{RESULTS}
\label{results}

\begin{figure}[h]
\centering
\includegraphics[scale=0.65]{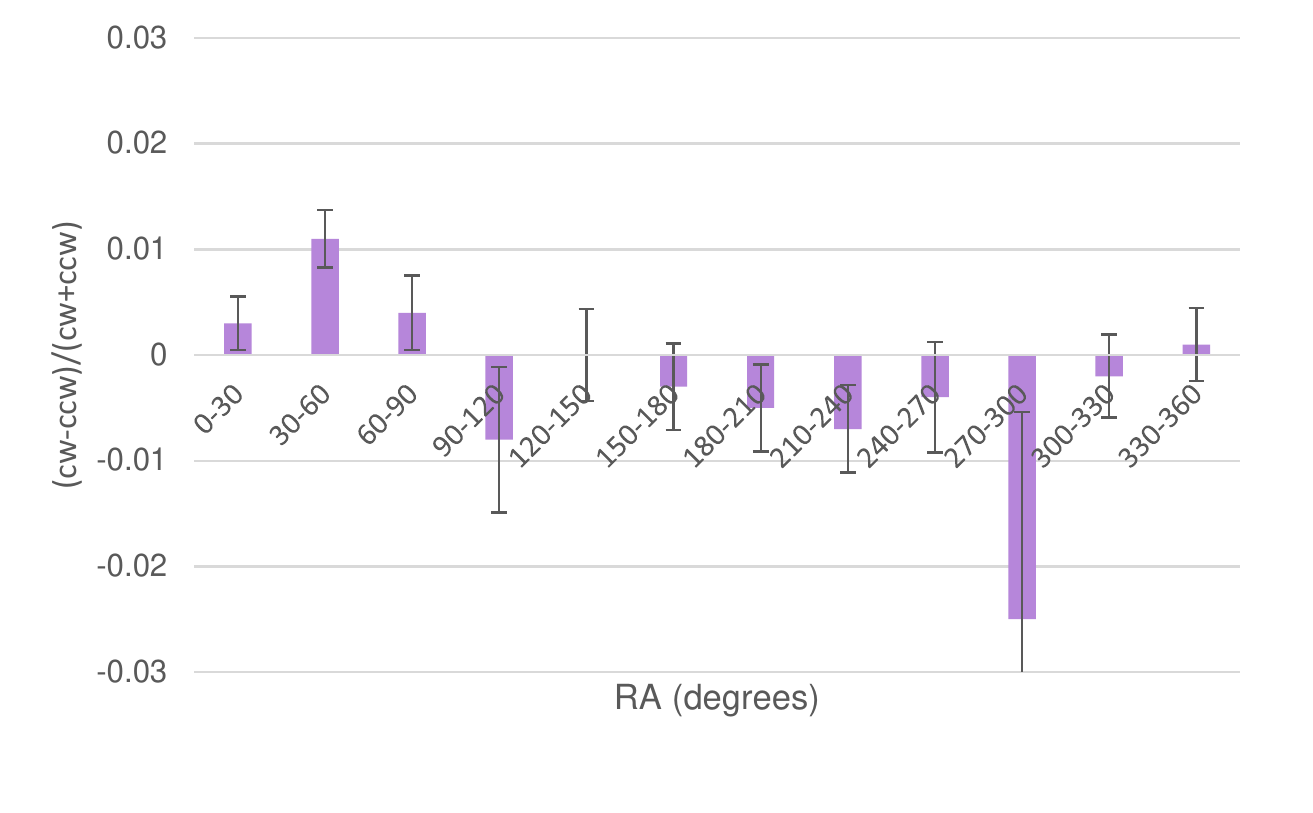}
\caption{Asymmetry in the distribution of the galaxies in different RA ranges. The graph shows inverse asymmetry in corresponding sky regions in opposite hemispheres.}
\label{by_ra}
\end{figure}

As a simple analysis, the asymmetry in each part of the sky was determined as $\frac{cw-ccw}{cw+ccw}$, where {\it cw} is the number of galaxies rotating clockwise, and {\it ccw} is the number of galaxies rotating counterclockwise in that part of the sky. The normal distribution standard error is $\frac{1}{\sqrt{N}}$, where {\it N} is the total number of galaxies in that sky section. Figure~\ref{by_ra} shows the asymmetry in different 30$^o$ RA slices. 

As the figure shows, the asymmetry in each RA slice agrees with the inverse asymmetry in the corresponding RA slice in the opposite hemisphere. For instance, the strongest asymmetry of 0.011 is observed in the RA range $(30^o-60^o)$. The uncorrected binomial probability to have such distribution by chance is $2\cdot10^{-5}$, and the Bonferroni corrected probability is $3\cdot10^{-4}$. That asymmetry corresponds to the strongest inverse asymmetry of -0.007, observed in the corresponding RA range in the opposite hemisphere $(210^o-240^o)$. The lowest asymmetry is observed in the RA range of $(120^o-150^o)$, and matches the low asymmetry in the corresponding sky region in the opposite hemisphere $(300^o-330^o)$. The same can be observed with all other sky regions, where the magnitude of the asymmetry in each sky region corresponds to the magnitude of the inverse asymmetry in the opposite sky region. The only exception is the RA range $(90^o-120^o)$, and the corresponding RA range $(270^o-300^o)$. But these RA ranges have by far the lowest number of galaxies 21,086 and 2,602, respectively, and therefore the error in these RA ranges is far larger compared to the other RA slices.

The symmetricity of the algorithm was verified by comparing the annotations of a certain set of galaxies to the annotations of the same set of galaxies such that the images are mirrored. Previous experiments with other telescopes showed that the annotation of the mirrored galaxy images are exactly the opposite compared to the original galaxy images \citep{shamir2017large,shamir2017colour,shamir2020patterns,shamir2021particles}. The fact that mirroring the galaxies provided inverse results is an empirical evidence that any error in the annotation algorithm affects both clockwise and counterclockwise galaxies in a similar fashion. % To verify the symmetricity of the analysis, the analysis was repeated such that each image was mirrored by converting the images to lossless TIFF format, and then using the ``flip'' command of ImageMagick. % Figure~\ref{by_ra_mirror} shows the results after mirroring the galaxy images. As expected, the results are inverse to the results with the original non-mirrored images. % invesre mirrored.

%\begin{figure}[h]
%\centering
%\includegraphics[scale=0.65]{by_ra_mirror.pdf}
%\caption{Asymmetry in the distribution of the galaxies in different RA ranges when analyzing the galaxies after mirroring each galaxy image.}
%\label{by_ra_mirror}
%\end{figure}

As Figure~\ref{by_ra} shows, the RA range $(150^o-330^o)$ shows higher population of counterclockwise galaxies, while the RA range $(0^o-150^o \cup  330^o-360^o)$ shows a higher number of galaxies that spin clockwise. The only exceptions are the RA slices with low population of galaxies. To verify the symmetricity of the analysis, the analysis was repeated such that each image was mirrored by converting the images to lossless TIFF format, and then using the ``flip'' command of ImageMagick. As expected, the results were exactly inverse to the results with the original non-mirrored images.

Table~\ref{hemispheres} shows the number of galaxies spinning in each direction in each hemisphere. As the table shows, the sky imaged by DESI Legacy Survey can be separated into two hemispheres such that one hemisphere has a higher number of galaxies that spin clockwise, and the opposite hemisphere has an excessive number of galaxies that spin counterclockwise. In both cases the asymmetry is statistically significant.

\begin{table*}
\caption{Number of clockwise and counterclockwise galaxies in opposite hemispheres. The P is the binomial distribution probability to have such difference or stronger by chance when assuming 0.5 probability for a galaxy to spin clockwise or counterclockwise.}
\label{hemispheres}
\centering
\begin{tabular}{lcccc}
\hline
\hline
Hemisphere       & \# cw galaxies & \# ccw galaxies  & $\frac{cw-ccw}{cw+ccw}$  & P \\
\hline
$(0^o-150^o \cup 330^o-360^o)$ &   264,707    &  262,559  &   0.004      &  0.0015  \\   
$(150^o-330^o)$                          &   139,719     & 140,913  &   -0.004    &  0.0121   \\
\hline
\hline
\end{tabular}
\end{table*}

The fact that the sky covered by the DESI Legacy Survey can be separated into two opposite parts such that one has a higher number of clockwise galaxies and the other has a higher number of counterclockwise galaxies indicates that the distribution of the spin directions of spiral galaxies as observed from Earth can form a large-scale axis. To examine the probability that the distribution of the spin directions of the galaxies exhibits a dipole axis, the same method used in  \citep{shamir2012handedness,shamir2019large,shamir2020patterns,shamir2020pasa,shamir2021particles} was applied. 

In summary, all galaxies in the dataset that spin clockwise were assigned with the value 1, and all galaxies that spin counterclockwise were assigned with the value -1. Then, for each possible integer $(\alpha,\delta)$ combination, the angular distance $\phi$ between each galaxy in the dataset and $(\alpha,\delta)$ was computed. The $\cos(\phi)$ of the galaxies were then fitted with $\chi^2$ statistics into $d\cdot|\cos(\phi)|$, where $d$ is the spin direction of the galaxy ($d$ can be either 1 or -1). The $\chi^2$ was computed 1000 times for each integer $(\alpha,\delta)$ combination such that in each time the galaxies were assigned with random spin directions,. The mean and standard deviation of the 1000 runs were computed for each possible integer $(\alpha,\delta)$. Then, the $\chi^2$ computed when $d$ was assigned the actual spin directions was compared to the mean and standard deviation of the $\chi^2$ computed with the random spin directions. The standard deviation was used to determine the $\sigma$ difference between the mean $\chi^2$ computed with the random galaxy spin directions and the $\chi^2$ computed with the actual spin directions. After computing the $\sigma$ difference of all $(\alpha,\delta)$ combinations, the location $(\alpha,\delta)$ of the most likely dipole axis can be determined by the $(\alpha,\delta)$ that has the highest $\sigma$ difference between the $\chi^2$ computed with the actual spin directions and the mean $\chi^2$ computed with the random spins. Detailed information about the analysis can be found in \citep{shamir2012handedness,shamir2019large,shamir2020pasa,shamir2020patterns,shamir2021particles}.

\begin{figure}[h]
\centering
\includegraphics[scale=0.15]{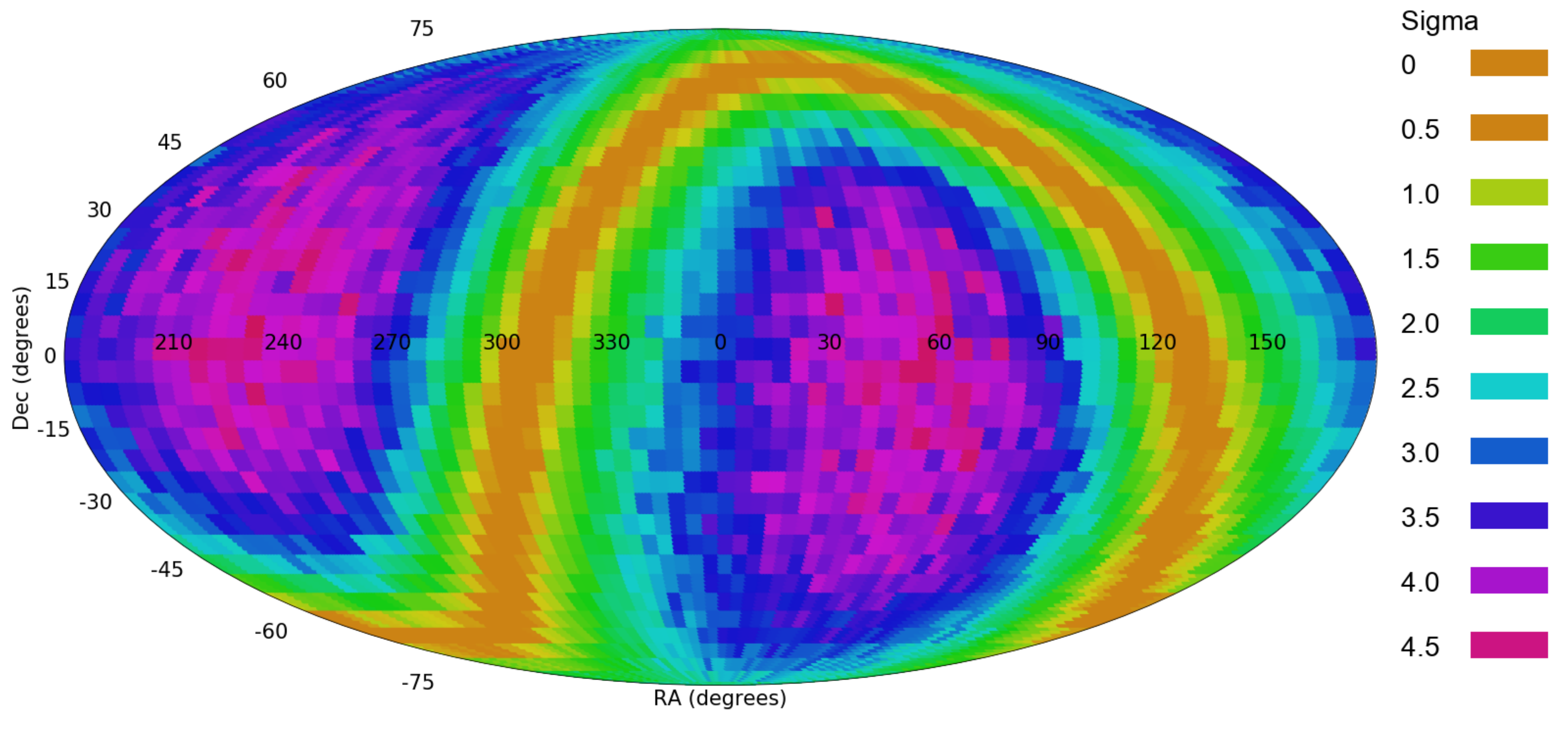}
\caption{The probability of a dipole axis in galaxy spin directions from every possible integer $(\alpha,\delta)$ combination.}
\label{decam_dipole}
\end{figure}

Figure~\ref{decam_dipole} shows the likelihood of a dipole axis in all $(\alpha,\delta)$ combinations. The most likely dipole axis was identified at $(\alpha=57^o,\delta=-10^o)$. The likelihood of the axis to be formed by chance is $4.66\sigma$. The 1$\sigma$ error for that axis is $(22^0,92^o)$ for the right ascension, and $(-39^o,56^o)$ for the declination. Interestingly, the CMB cold spot is at $(\alpha=49,\delta=-19^o)$, very close to the direction of the most probable dipole axis.

\begin{figure}[h!]
\centering
\includegraphics[scale=0.15]{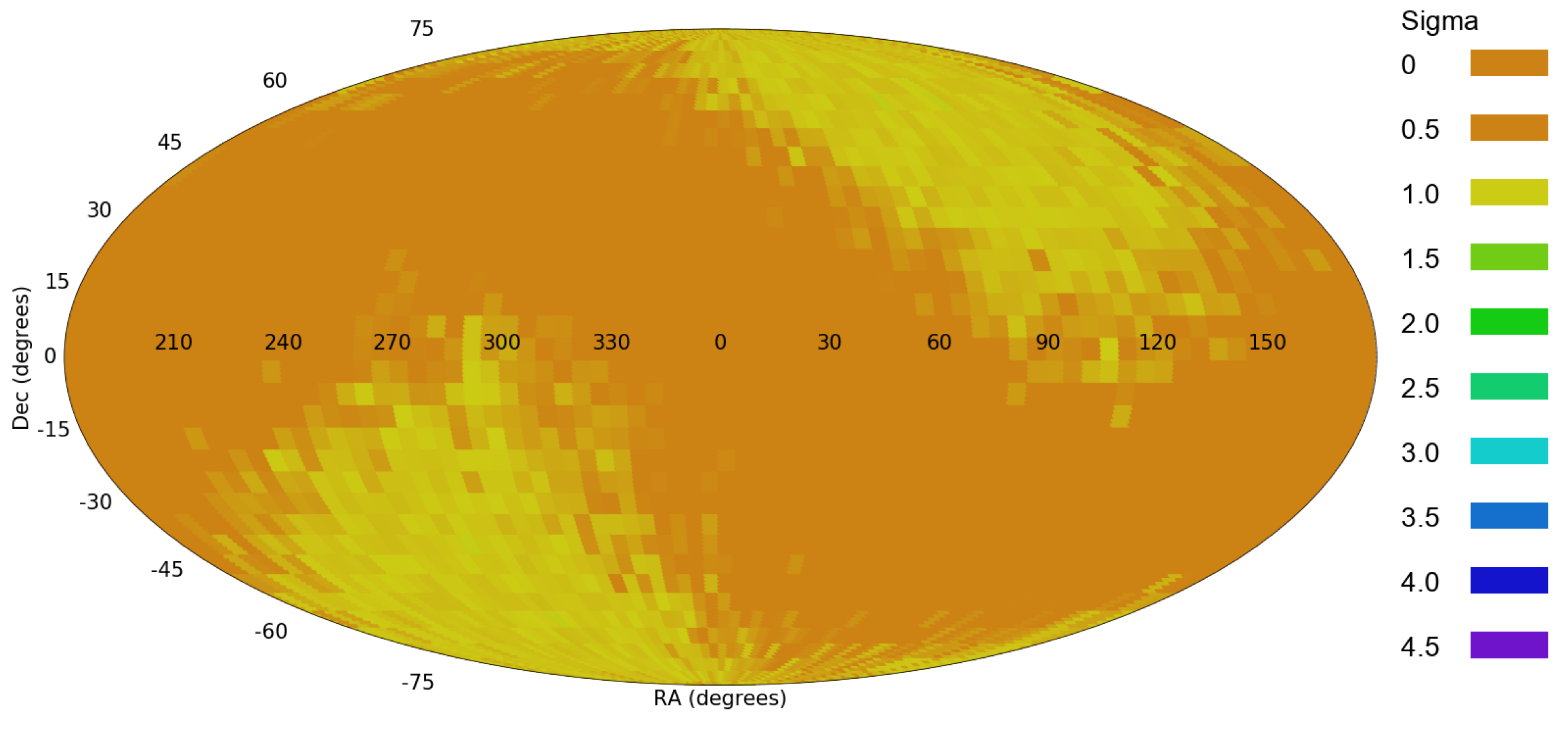}
\caption{The probability of a dipole axis in the galaxy spin directions when the galaxies are assigned with random spin directions.}
\label{decam_dipole_random}
\end{figure}

Assigning the galaxies with random spin directions provides a maximum likelihood of a dipole of 0.91$\sigma$. The likelihood of a dipole axis formed by the spin directions of the galaxies in all integer $(\alpha,\delta)$ combinations is shown in Figure~\ref{decam_dipole_random}. As expected, galaxies with random spin directions do not form a statistically significant pattern.

\begin{figure}[h!]
\centering
\includegraphics[scale=0.15]{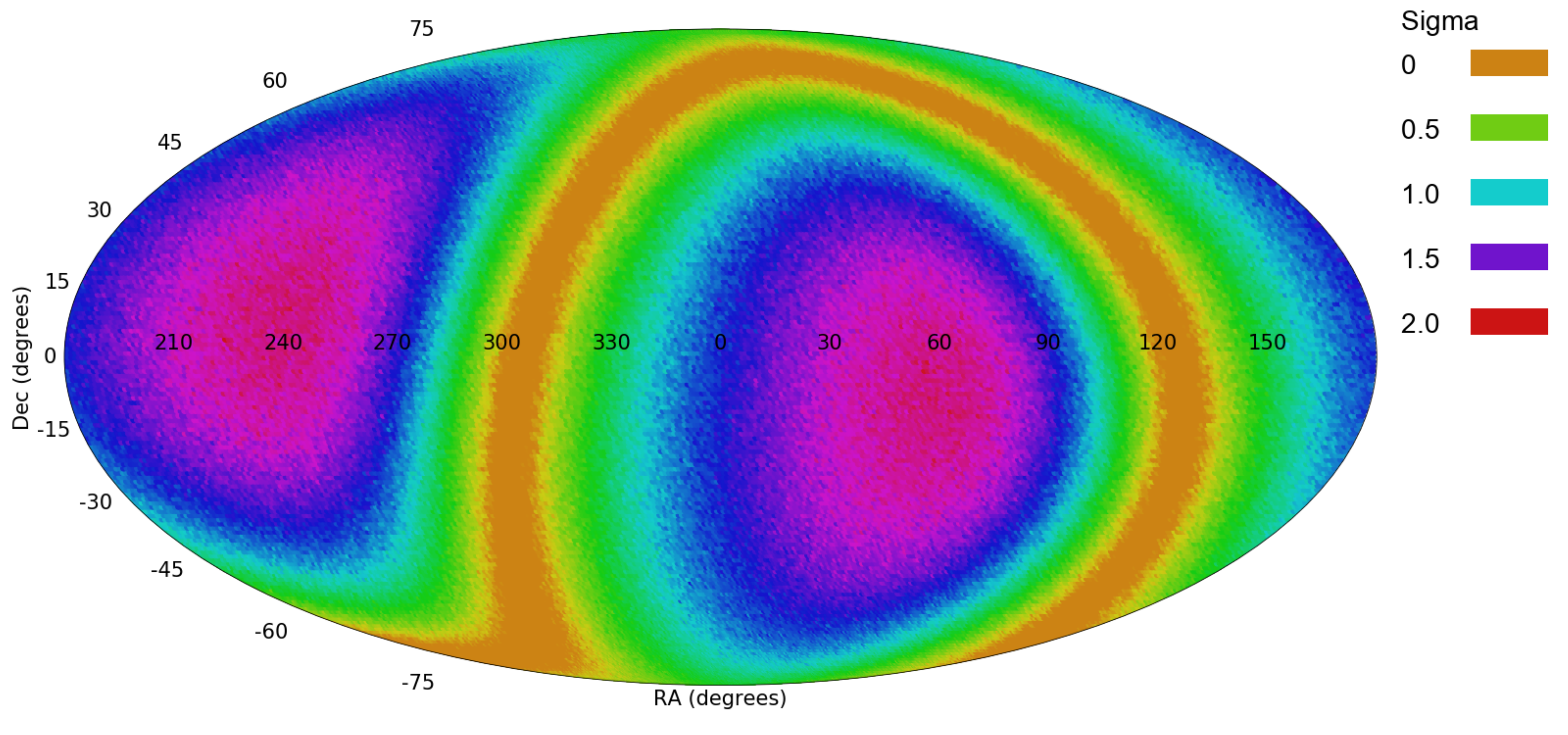}
\caption{The probability of a dipole axis formed by the galaxy spin directions of $\sim3.3\cdot10^4$ Pan-STARRS galaxies \citep{shamir2020patterns}. The profile is very similar to the profile formed by the galaxy spin directions of the galaxies imaged by the DESI Legacy Survey shown in Figure~\ref{decam_dipole}.}
\label{PanSTARRS_DR14_dipole}
\end{figure}

The results observed with the DESI Legacy Survey, in which most galaxies are in the Southern hemisphere, can be compared to previous results using galaxies mostly from the Northern hemisphere, namely Pan-STARRS, SDSS, and HST. Figure~\ref{PanSTARRS_DR14_dipole} shows the results of a previous experiment \citep{shamir2020patterns} of fitting the spin directions of 3.3$\cdot10^3$ Pan-STARRS galaxies into cosine dependence from all possible integer $(\alpha,\delta)$ combinations. The specific details of the experiment are described in \citep{shamir2020patterns}. The most likely dipole axis in Pan-STARRS galaxies is identified at $(\alpha=47^o,\delta=-1^o)$, with statistical significance of 1.87$\sigma$ \citep{shamir2020patterns}. That location is nearly identical to the most probable axis identified in the DESI Legacy Survey galaxies, shown in Figure~\ref{decam_dipole}.

\begin{figure}[h]
\centering
\includegraphics[scale=0.15]{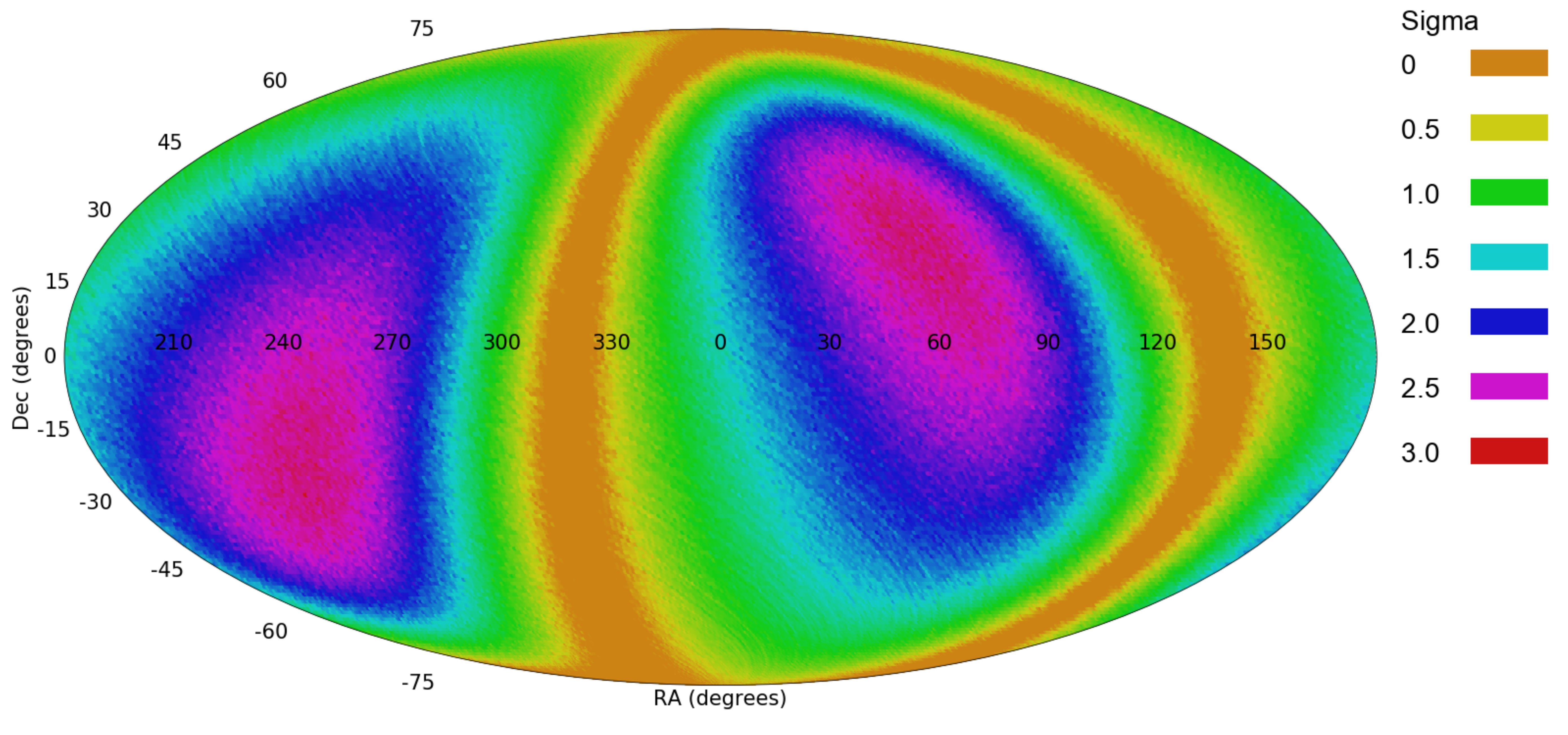}
\caption{The probability of a dipole axis in different $(\alpha,\delta)$ when using $\sim3.3\cdot10^4$ SDSS galaxies \citep{shamir2020patterns}.}
\label{DR14_PanSTARRS_dipole}
\end{figure}

The previous analysis of the distribution of the spin directions of the galaxies imaged by Pan-STARRS was compared to the previous analysis of 38,998 SDSS galaxies with spectra that their redshifts distribute in a similar manner to the redshift distribution of the Pan-STARRS galaxies. Full details about that experiment are described in \citep{shamir2020patterns}. Figure~\ref{DR14_PanSTARRS_dipole} shows the probability of a dipole axis in the distribution of spin directions of the SDSS DR14 spiral galaxies \citep{shamir2020patterns}. The results show a most likely dipole axis at $(\alpha=49^o, \delta=21^o)$, with statistical signal of 2.05$\sigma$ \citep{shamir2020patterns}. That position is also well within the 1$\sigma$ error from the most likely dipole axis identified in the DESI Legacy Survey. The difference in the RA of the two axes is just 7$^o$.

\begin{figure}[h]
\centering
\includegraphics[scale=0.15]{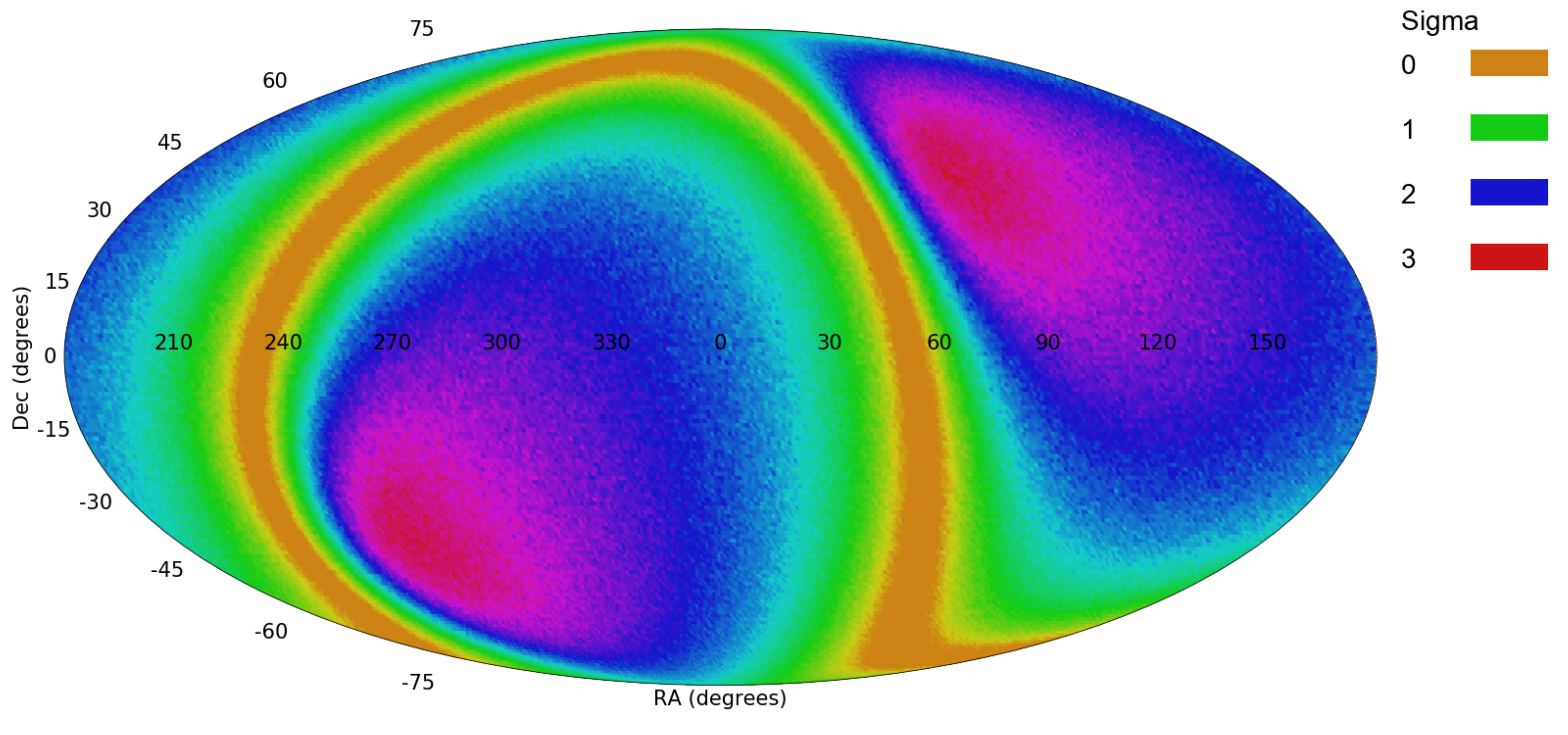}
\caption{The profile of probabilities of a dipole axis in the spin directions of HST galaxies \citep{shamir2020pasa}.}
\label{hst_dipole}
\end{figure}

The results can also be compared to the previous analysis of using $\sim8.7\cdot10^3$ Hubble Space Telescope (HST) galaxies classified manually by their spin directions \citep{shamir2020pasa}. Figure~\ref{hst_dipole} shows the probability of a dipole axis in all possible integer $(\alpha,\delta)$ combinations. The most likely dipole axis exhibited by the HST galaxies was identified at $(\alpha=78^o, \delta=47^o)$, with statistical significance of 2.8$\sigma$ as described in \citep{shamir2020pasa}. The 1$\sigma$ error for that axis is $(58^o, 184^o)$ for the right ascension, and $(6^o, 73^o)$ for the declination. While that axis is not as close to the axis of the DESI Legacy Survey data as the axes observed in Pan-STARRS and SDSS data, the right ascension is still relatively close, and the 1$\sigma$ error of that axis is still within the 1$\sigma$ error of the most probable dipole axis computed from the DESI Legacy Survey. It should also be noted that HST has a lower number of galaxies compared to the three other telescopes, and the mean redshift of the HST galaxies of 0.58 is far higher than all other sky surveys.

\begin{figure}[h]
\centering
\includegraphics[scale=0.15]{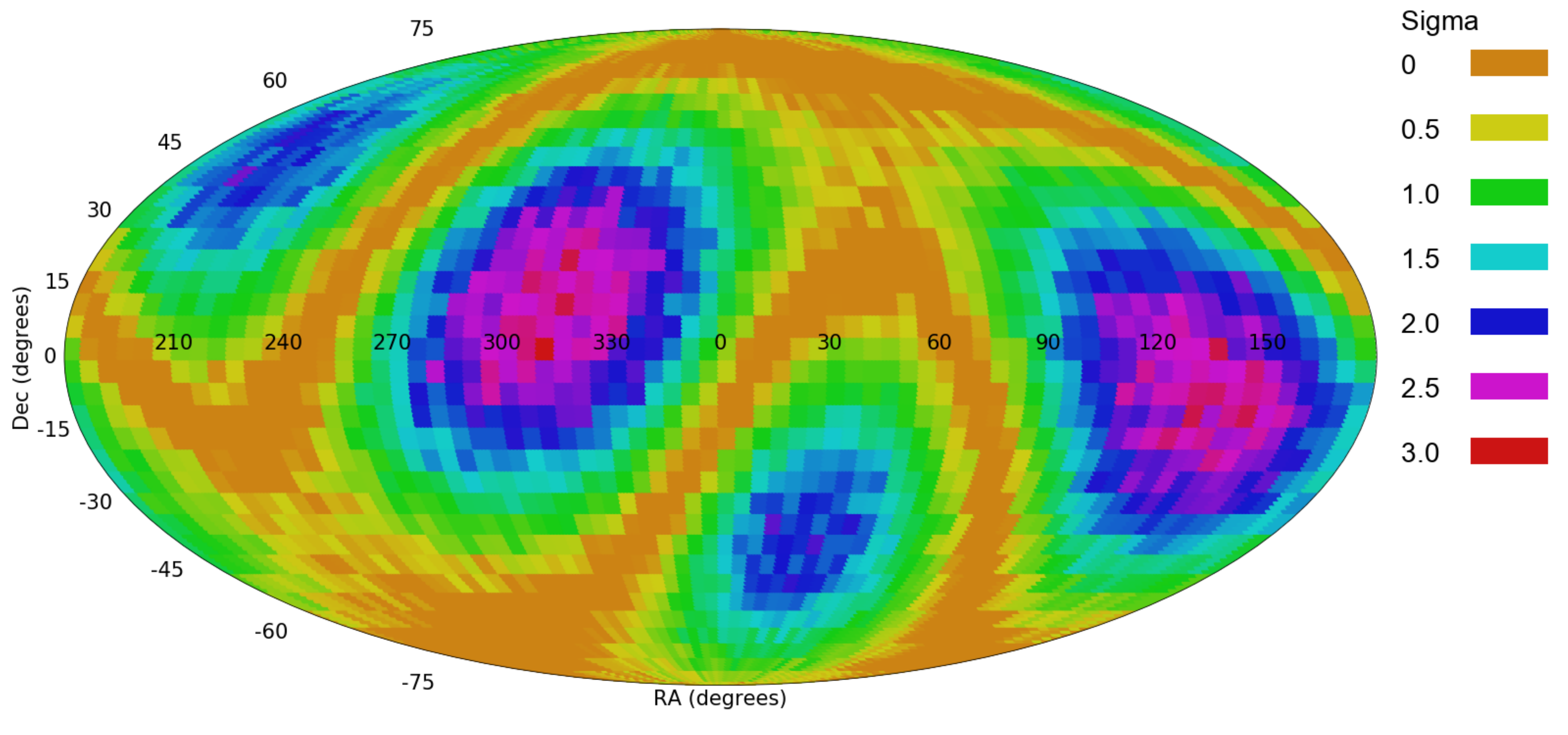}
\caption{The profile of probabilities of a quadrupole axis in the spin directions of DESI Legacy Survey galaxies.}
\label{decam_quad}
\end{figure}

In addition to a dipole axis, an attempt was also made to fit the galaxy spin directions to quadrupole alignment. That was done by $\chi^2$ fitting of the $\cos(2\phi)$ into $d\cdot|\cos(2\phi)|$ \citep{shamir2019large,shamir2020patterns}. Figure~\ref{decam_quad} shows the likelihood of a quadrupole axis from different combinations of $(\alpha,\delta)$. The most probable quadrupole axis show one axis at $(\alpha=312^o, \delta=1^o)$, and another axis at $(\alpha=27^o, \delta=-32^o)$, with statistical significance of 3.069$\sigma$ and $1.87\sigma$, respectively. The statistical significance of the quadrupole alignment is therefore lower than the statistical significance of the dipole alignment.

\begin{figure}[h]
\centering
\includegraphics[scale=0.15]{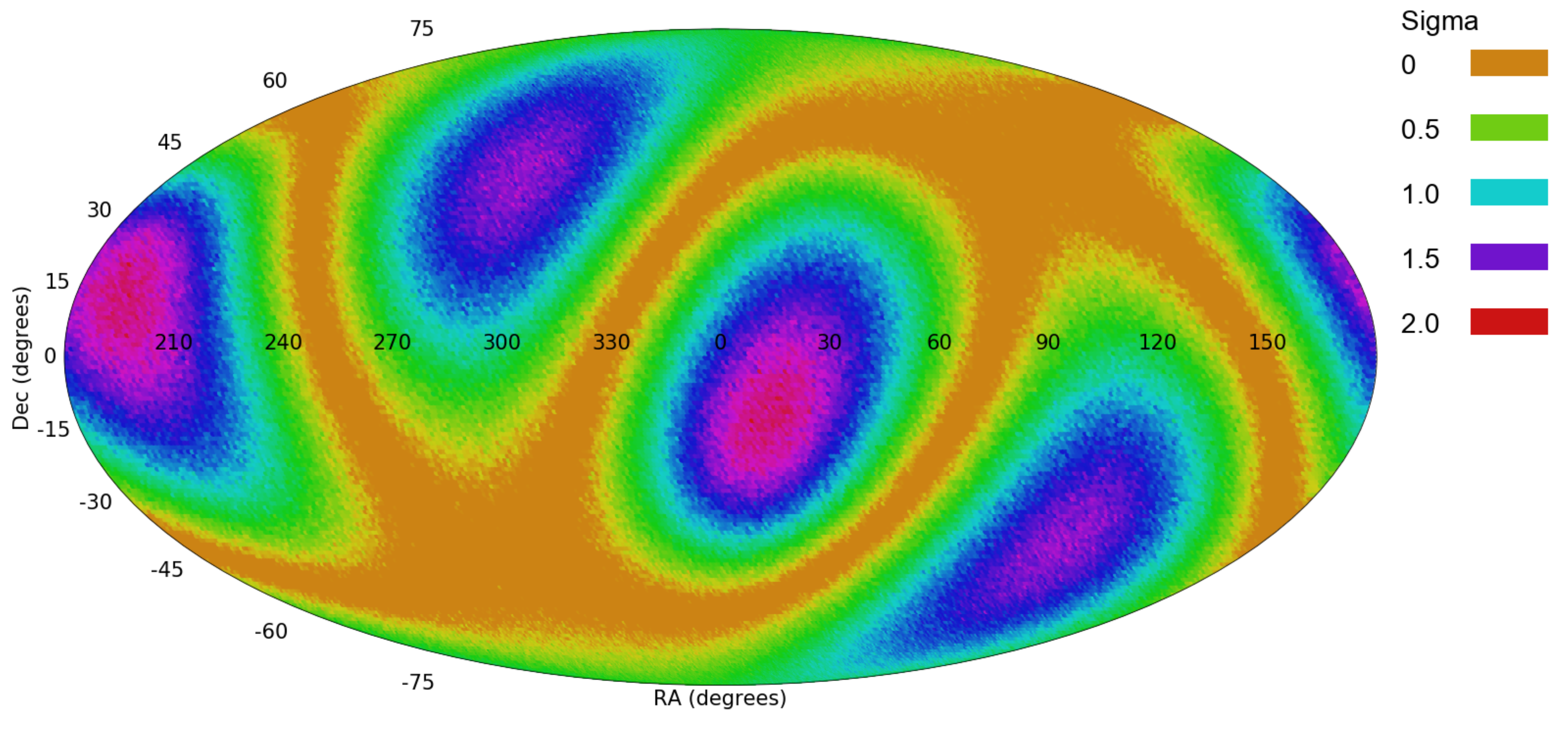}
\caption{The profile of a quadrupole in Pan-STARRS \citep{shamir2020patterns}.}
\label{PanSTARRS_DR14_quad}
\end{figure}

\begin{figure}[h]
\centering
\includegraphics[scale=0.5]{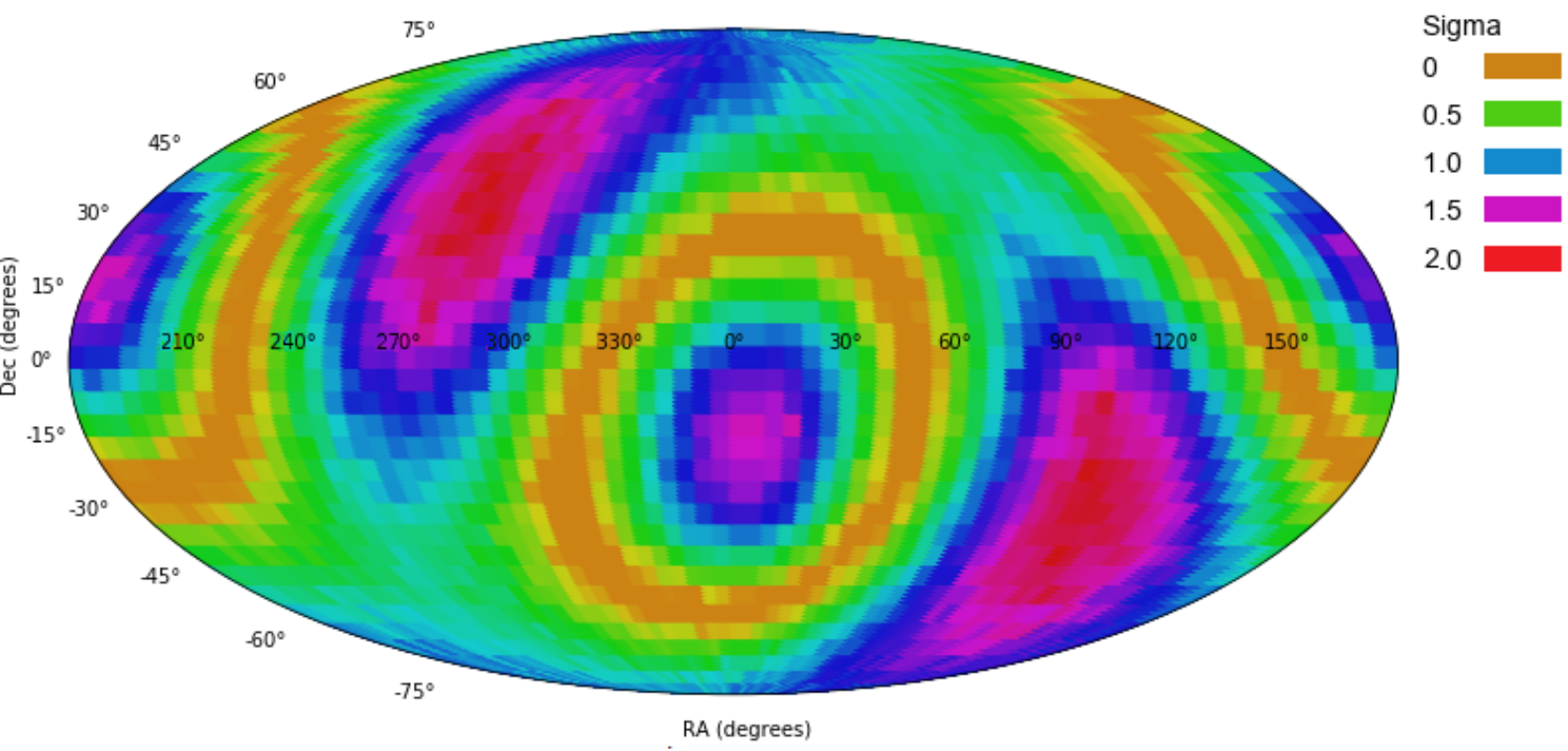}
\caption{The profile of a quadrupole in SDSS \citep{shamir2020patterns}.}
\label{DR14_PanSTARRS_quad}
\end{figure}

These axes can be compared to the quadrupole alignment of the Pan-STARRS and SDSS galaxies, shown in Figures~\ref{PanSTARRS_DR14_quad} and~\ref{DR14_PanSTARRS_quad}, respectively. The most probable axes are $(\alpha=17^o, \delta=-2^o)$ in Pan-STARRS, and $(\alpha=7^o, \delta=4^o)$ in SDSS \citep{shamir2020patterns}. Fitting the galaxies into quadrupole alignment shows somewhat larger differences between the different telescopes compared to the strong agreement between the dipole axes. However, the difference in the RA is still relatively small, and is 10$^o$ difference in Pan-STARRS, and 20$^o$ in SDSS.

\section{CONCLUSION}
\label{conclusion}

Several previous observations have shown the possibility of certain asymmetry between galaxies with opposite spin directions, and large-scale patterns that the asymmetry might exhibit as observed from Earth \citep{longo2011detection,shamir2012handedness,shamir2013color,shamir2016asymmetry,shamir2019large,shamir2020patterns,lee2019mysterious,lee2019galaxy,shamir2020pasa,shamir2021particles}. While these observations show good agreement between telescopes, previous experiments were based on galaxies imaged mostly in the Northern hemisphere. The analysis shown in this paper is based on a dataset in which the majority of the galaxies are from the Southern hemisphere, which is also far larger than any previous dataset used for this purpose before.

Simple analysis of asymmetry in different RA slices shows good agreement between the asymmetry in corresponding RA slices in opposite hemispheres, such that a higher number of clockwise galaxies in one part of the sky corresponds to a similarly higher number of counterclockwise galaxies in the corresponding part of the sky in the opposite hemisphere. Analysis of a dipole axis shows good agreement with previous data from Pan-STARRS, SDSS, and HST. The data is available in \url{http://people.cs.ksu.edu/~lshamir/data/assym_desi/}.

It is very difficult to think of an error that would exhibit itself in the form of such asymmetry. The classification algorithm is mathematically symmetric, and was tested empirically to show inverse results when the images are mirrored. Even if the annotation algorithm had an error, that error should have been consistent throughout the sky, rather than being flipped in opposite hemispheres. Obviously, no change in the algorithm or code was done while the analysis was performed, and all galaxies were annotated by the exact same algorithm and code, the exact same computer, and the exact same processor.

The algorithm used to annotate the galaxies is fully symmetric, and works by clear mathematically defined rules. Therefore, an error in the galaxy annotation should affect both clockwise and counterclockwise galaxies similarly, and therefore cannot lead to asymmetry. That is also true for artefacts or bad data that might be relatively rare but still exist in large databases such as the DESI Legacy Survey. A detailed theoretical and empirical analysis of the effect of error in the galaxy annotation was done in \citep{shamir2021particles}. In summary, the asymmetry between clockwise and counterclockwise galaxies {\it A} can be defined as $A=\frac{(N_{cw}+E_{cw})-(N_{ccw}+E_{ccw})}{N_{cw}+E_{cw}+N_{ccw}+E_{ccw}}$, where $N_{cw}$ is the number of clockwise galaxies correctly annotated as clockwise, $N_{ccw}$ is the number of counterclockwise galaxies correctly annotated as counterclockwise, $E_{cw}$ is the number of counterclockwise galaxies incorrectly classified as clockwise galaxies, and $E_{ccw}$ is the number of clockwise galaxies incorrectly classified as counterclockwise galaxies.

If the galaxy classification algorithm is symmetric, $E_{cw}$ should be roughly the same as $E_{ccw}$, and therefore the asymmetry can be defined by $A=\frac{N_{cw}-N_{ccw}}{N_{cw}+E_{cw}+N_{ccw}+E_{ccw}}$. Since $E_{cw}$ and $E_{ccw}$ cannot be negative values, a higher number of misclassified galaxies is expected to make {\it A} smaller. Therefore, an error in the annotation algorithm is expected to make the asymmetry lower, and cannot lead to a statistically significant {\it A} in a population of evenly distributed galaxies. The analysis agrees with empirical experiments done by adding an artificial error to the galaxy annotation algorithm, showing that such error weakens the signal \citep{shamir2021particles}. Therefore, if the algorithm is symmetric, error in the annotations is expected to weaken the signal, and cannot lead to statistically significant asymmetry if the real distribution of the galaxies is random. In some rare cases the actual spin of a galaxy can be different from the curves of the arms. However, these cases should be distributed equally between clockwise and counterclockwise galaxies, and therefore are not expected to lead to signal in the distribution of spin directions.

Cosmic variance also cannot explain such asymmetry. The difference between the number of clockwise and counterclockwise galaxies is a relative measurement, made of two measurements made in the same field. The number of clockwise galaxies is one measurement, while the number of counterclockwise spiral galaxies is another measurement made from the same field. Any effect that can impact the number of clockwise galaxies identified in the field is expected to make the same impact on the number of counterclockwise galaxies.

As mentioned in Section~\ref{data}, in some cases photometric objects can be part of the same galaxy, and therefore any object that has another object within less than 0.01$^o$ was removed from the dataset to ensure that no galaxy is represented more than once. Previously, \cite{iye2020spin} reported on photometric objects that are part of the same galaxies in the dataset of \citep{shamir2017photometric}. If that dataset was used for the purpose of profiling asymmetry in the population of clockwise and counterclockwise galaxies, photometric objects that are part of the same galaxy become duplicate objects. However, the dataset of \citep{shamir2017photometric} was used to analyze the possible link between photometry and spin direction, and not for analyzing the large-scale distribution of clockwise and counterclockwise galaxies as done here or in previous studies \citep{shamir2012handedness,shamir2019large,shamir2020pasa,shamir2020patterns,shamir2020large,shamir2021particles}. That is, if the dataset of \citep{shamir2017photometric} was used to identify a dipole axis in the clockwise/counterclockwise distribution, the photometric objects that are part of the same galaxies would have been duplicate objects. However, the \citep{shamir2017photometric} was designed for a different purpose, and was not used for identifying a dipole axis in the clockwise/counterclockwise galaxy distribution.

It is important to mention that unlike the analysis applied here, in which the location of each galaxy is determined by its RA and Dec, \cite{iye2020spin} applied a 3D analysis, which required each galaxy to have its RA, Dec, and redshift. Because the vast majority of the galaxies used in \citep{shamir2017photometric} do not have spectra, \cite{iye2020spin} used the photometric redshifts taken from the photometric redshift catalog of \citep{paul2018}. The error of the photometric redshift in that catalog is $\sim$18.5\%, which is far greater than the expected asymmetry. Also, because photometric objects that are part of the same galaxy can be assigned with different photometric redshifts, such analysis with the photometric redshift can lead to biased results that are not of astronomical origin. In summary, using the photometric redshift for identifying subtle asymmetries in the large-scale structure might be limited by the relatively large error and the possible systematic bias of the photometric redshift, and might not provide sound evidence for the existence or inexistence of such cosmological-scale asymmetries.

Another difference between the study of \cite{iye2020spin} and the work done in previous papers \citep{shamir2012handedness,shamir2020patterns,shamir2020pasa} is that \cite{iye2020spin} reported on random distribution in a subset of the data, limited to $z_{phot}<0.1$. As shown in \citep{shamir2020patterns}, no statistically significant asymmetry is expected in that redshift range, and in fact all previous attempts to limit the redshift to any value below 0.15 showed random distribution \citep{shamir2020patterns}. An experiment of identifying a dipole axis when limiting the galaxies to $z<0.15$ showed statistical signal lower than 2$\sigma$ \citep{shamir2020patterns}, and the distribution in different parts of the sky showed low P values for the lower redshift ranges, as shown in Tables, 3, 5, 6, and 7 in \citep{shamir2020patterns}.

A dataset from the same telescope (SDSS) and the same limiting magnitude (g$<$19) that does not contain duplicate objects was used for the analysis of spin direction asymmetry and a possible dipole axis. The analysis showed that when not using the photometric redshift and when not limiting the galaxies to $z_{phot}<0.1$, the patterns are statistically significant at $\sim2.6\sigma$ \citep{shamir2021particles}.

% The (Iye et al, 2021) paper also reports its main findings when using a subset of z_{phot}<0.1. Not only that I never claimed for a dipole axis in that redhisft range, I showed explicitly in (Shamir, 2020b) that when limitting to z<0.15 the asymmetry is not statistically significant. (Shamir, 2020b) shows an experiment when limiting z<0.15, and Tables 3, 5, 6, and 7 in that paper also show weak P values for the lower redshifts. The random distribution at z_{phot}<0.1 is therefore in full agreement with the random distribtuion at z<0.15 and other lower redshift ranges shown in (Shamir, 2020b). I argue that a valid experiment can only be done when using the entire dataset. 

In any case, duplicate objects are expected to be distributed evenly between galaxies that spin clockwise and galaxies that spin counterclockwise. Empirical experiments by adding artificial duplicate objects to randomly distributed data showed that duplicate objects do not have substantial impact on the results, until an extremely large number of duplicate objects is present. The results show that even when duplicating each object five times, the asymmetry signal is still below 2$\sigma$ if the original data is random \citep{shamir2021particles}.

The contention that alignment in galaxy spin directions is related to the large-scale structure was also proposed with smaller scale experiments, in which correlation between spin directions of spiral galaxies were identified even when the galaxies were too far to have gravitational interactions \citep{lee2019mysterious}. Unless assuming modified Newtonian dynamics (MOND) gravity models that support longer gravitational span \citep{amendola2020measuring}, these findings might conflict with standard cosmology. Alignment of the position angle of radio galaxies also showed large-scale consistency of angular momentum \citep{taylor2016alignments}. These observations have been aligned with observations made with datasets such as the TIFR GMRT Sky Survey (TGSS) and the Faint Images of the Radio Sky at Twenty-centimetres (FIRST), showing large-scale alignment of radio galaxies \citep{contigiani2017radio,panwar2020alignment}.

% These observations also conflict with the standard cosmology, unless the assumption of Newtonian gravity is expanded into modified Newtonian dynamics (MOND) models that can explain longer gravitational span \citep{amendola2020measuring}, while also explaining other anomalies in the context of the standard model such as the Keenan–Barger–Cowie (KBC) void \citep{haslbauer2020kbc}.

Large-scale anisotropy has been reported also by using several other probes such as the cosmic microwave background \citep{eriksen2004asymmetries}, frequency of galaxy morphology types \citep{javanmardi2017anisotropy}, short gamma ray bursts \citep{meszaros2019oppositeness}, Ia supernova \citep{javanmardi2015probing,lin2016significance}, LX-T scaling \citep{migkas2020probing}, and quasars \citep{quasars}. The cosmic microwave background also shows the possibility of the existence of a cold spot \citep{cruz655non,mackenzie2017evidence,farhang2021cmb}, which makes another indication of the possibility of cosmological-scale anisotropy. The location of the most probable location of the dipole axis shown in this paper is very close to the location of the CMB cold spot.

The possible anisotropy observed in the cosmic microwave background \citep{cline2003does,gordon2004low,zhe2015quadrupole} has led to cosmological theories that challenge the standard cosmology models. These theories include primordial anisotropic vacuum pressure \citep{rodrigues2008anisotropic}, double inflation \citep{feng2003double}, contraction prior to inflation \citep{piao2004suppressing}, moving dark energy \citep{jimenez2007cosmology}, multiple vacua \citep{piao2005possible}, and spinor-driven inflation \citep{bohmer2008cmb}. Understanding and profiling the possible cosmological-scale anisotropy can provide important information leading to additional theories that shift from the existing standard models.

% These measurement are also close to the newer space-based measurements of the dipole direction at $\alpha=157^o, \delta=0^o$.

%The alignment in spin directions between galaxies too far from each other to have a gravitational link can also be considered an indication of modification of the Newtonian dynamics assumed in the standard cosmology. Such gravitational dynamics can also be aligned with the Hubble tension by means of our location within a particularly wide and deep supervoid that is not expected in $\Lambda$CDM. While not expected by the standard model, such a void could arise in a Milgromian cosmology. That is, the expansion rate history, cosmic microwave background anisotropies, and primordial light element abundances would remain the same as in $\Lambda$CDM, but structure formation would be enhanced due to the modified gravity law. Enhanced structure formation compared to $\Lambda$CDM is also suggested by the El Gordo (ACT-CL J0102-4915) massive galaxy cluster, which has observed properties that conflict with $\Lambda$CDM at probability of 6.16 $\sigma$ \citep{asencio2020massive}. Other tensions between $\Lambda$CDM focus on smaller scale observations, normally at scales smaller than 1 Mpc \citep{del2017small,bullock2017small}. For instance, the satellite plane and the number of dwarf satellite galaxies in the Local Group shifts from what $\Lambda$CDM predicts, and galaxies dominated by dark matter are not as dense as can be deduced from $\Lambda$CDM. Dwarf galaxies can also have a much flatter core than predicted by $\Lambda$CDM \citep{pozo2020detection}.

A large-scale axis can be related also to other cosmological models such as rotating universe \citep{godel1949example,ozsvath1962finite,ozsvath2001approaches,sivaram2012primordial,chechin2016rotation}, or ellipsoidal universe \citep{campanelli2006ellipsoidal,campanelli2007cosmic,gruppuso2007complete,campanelli2011cosmic,cea2014ellipsoidal}, where a large-scale asymmetry axis is assumed, and has also been associated with the axis formed by CMB anisotropy \citep{campanelli2007cosmic}.

Cosmological theories such as holographic big bang \citep{pourhasan2014out,altamirano2017cosmological} can also be related to a cosmological-scale axis. The existence of such axis is also aligned with the black hole cosmology theory \citep{pathria1972universe,easson2001universe,chakrabarty2020toy}, providing an explanation to cosmic inflation that does not involve dark energy. The spin of black holes is originated from the spin of the stars from which they were created \citep{mcclintock2006spin}. Since a black hole spins, a cosmological-scale axis is expected to exist in the universe hosted by it. If the Universe was formed in a black hole, the Universe should have a preferred direction inherited from the spin direction of the black hole hosting it \citep{poplawski2010cosmology,seshavatharam2010physics,seshavatharam2020light}, and therefore an axis \citep{seshavatharam2020integrated}. Such black hole universe might not be aligned with the cosmological principle \citep{stuckey1994observable}.

% https://en.wikipedia.org/wiki/Black_hole_cosmology
% axis of evil

% A universe born from a rotating black hole should inherit its preferred direction, related to the axis of rotation.   
% Friedmann–Lemaître–Robertson–Walker
% https://www.sciencedirect.com/science/article/pii/S0370269310011561?via%3Dihub

% or alternatively could be caused by large-scale filamentary structures
% in the cosmic web giving rise to preferential merger directions (e.g. Chiaberge et al. 2015; Croton et al. 2006), although the observational evidence for both remains incomplete. Taylor & Jagannathan (2016) found a local alignment of radio galaxies in the ELAIS N1 field on scales using observations from the Giant Metrewave Radio Telescope (GMRT) at 610 MHz. Local alignments were also tentatively confirmed by Contigiani et al. (2017) who reported marginal evidence (¡ 2f) of local alignment among radio sources using a much larger sample of radio galaxies from the FIRST survey, catalogued by the radio galaxy zoo project. A similar local alignment was also reported by Panwar et al. (2020)

The ability to analyze and profile non-random distribution of the spin directions of galaxies is a relatively new research topic that became possible due to the availability of digital sky survey. These studies were not possible in the pre-information era. As the evidence for the existence of such axis are accumulating, it is clear that further research will be required to fully understand the nature of the possible non-random structures formed by the distribution of spin directions of galaxies.

\section*{ACKNOWLEDGMENTS}

I would like to thank the anonymous reviewer for the insightful comments that helped to improve the manuscript. This study was supported in part by NSF grants AST-1903823 and IIS-1546079. I would like to thank Ethan Nguyen for downloading and organizing the data from DESI Legacy Survey.

The Legacy Surveys consist of three individual and complementary projects: the Dark Energy Camera Legacy Survey (DECaLS; Proposal ID \#2014B-0404; PIs: David Schlegel and Arjun Dey), the Beijing-Arizona Sky Survey (BASS; NOAO Prop. ID \#2015A-0801; PIs: Zhou Xu and Xiaohui Fan), and the Mayall z-band Legacy Survey (MzLS; Prop. ID \#2016A-0453; PI: Arjun Dey). DECaLS, BASS and MzLS together include data obtained, respectively, at the Blanco telescope, Cerro Tololo Inter-American Observatory, NSF’s NOIRLab; the Bok telescope, Steward Observatory, University of Arizona; and the Mayall telescope, Kitt Peak National Observatory, NOIRLab. The Legacy Surveys project is honored to be permitted to conduct astronomical research on Iolkam Du’ag (Kitt Peak), a mountain with particular significance to the Tohono O’odham Nation.

NOIRLab is operated by the Association of Universities for Research in Astronomy (AURA) under a cooperative agreement with the National Science Foundation.

This project used data obtained with the Dark Energy Camera (DECam), which was constructed by the Dark Energy Survey (DES) collaboration. Funding for the DES Projects has been provided by the U.S. Department of Energy, the U.S. National Science Foundation, the Ministry of Science and Education of Spain, the Science and Technology Facilities Council of the United Kingdom, the Higher Education Funding Council for England, the National Center for Supercomputing Applications at the University of Illinois at Urbana-Champaign, the Kavli Institute of Cosmological Physics at the University of Chicago, Center for Cosmology and Astro-Particle Physics at the Ohio State University, the Mitchell Institute for Fundamental Physics and Astronomy at Texas A\&M University, Financiadora de Estudos e Projetos, Fundacao Carlos Chagas Filho de Amparo, Financiadora de Estudos e Projetos, Fundacao Carlos Chagas Filho de Amparo a Pesquisa do Estado do Rio de Janeiro, Conselho Nacional de Desenvolvimento Cientifico e Tecnologico and the Ministerio da Ciencia, Tecnologia e Inovacao, the Deutsche Forschungsgemeinschaft and the Collaborating Institutions in the Dark Energy Survey. The Collaborating Institutions are Argonne National Laboratory, the University of California at Santa Cruz, the University of Cambridge, Centro de Investigaciones Energeticas, Medioambientales y Tecnologicas-Madrid, the University of Chicago, University College London, the DES-Brazil Consortium, the University of Edinburgh, the Eidgenossische Technische Hochschule (ETH) Zurich, Fermi National Accelerator Laboratory, the University of Illinois at Urbana-Champaign, the Institut de Ciencies de l’Espai (IEEC/CSIC), the Institut de Fisica d’Altes Energies, Lawrence Berkeley National Laboratory, the Ludwig Maximilians Universitat Munchen and the associated Excellence Cluster Universe, the University of Michigan, NSF’s NOIRLab, the University of Nottingham, the Ohio State University, the University of Pennsylvania, the University of Portsmouth, SLAC National Accelerator Laboratory, Stanford University, the University of Sussex, and Texas A\&M University.

BASS is a key project of the Telescope Access Program (TAP), which has been funded by the National Astronomical Observatories of China, the Chinese Academy of Sciences (the Strategic Priority Research Program “The Emergence of Cosmological Structures” Grant \# XDB09000000), and the Special Fund for Astronomy from the Ministry of Finance. The BASS is also supported by the External Cooperation Program of Chinese Academy of Sciences (Grant \# 114A11KYSB20160057), and Chinese National Natural Science Foundation (Grant \# 11433005).

The Legacy Survey team makes use of data products from the Near-Earth Object Wide-field Infrared Survey Explorer (NEOWISE), which is a project of the Jet Propulsion Laboratory/California Institute of Technology. NEOWISE is funded by the National Aeronautics and Space Administration.

The Legacy Surveys imaging of the DESI footprint is supported by the Director, Office of Science, Office of High Energy Physics of the U.S. Department of Energy under Contract No. DE-AC02-05CH1123, by the National Energy Research Scientific Computing Center, a DOE Office of Science User Facility under the same contract; and by the U.S. National Science Foundation, Division of Astronomical Sciences under Contract No. AST-0950945 to NOAO.

\bibliographystyle{apalike}
\bibliography{arxiv}

\end{document}